\documentclass{aa}
\usepackage{graphicx} 
\usepackage{txfonts}
\usepackage{physics}
\usepackage{xspace}
\usepackage{xcolor}
\usepackage{tabularx}
\usepackage{amsmath}
\usepackage{booktabs}
\usepackage{hyperref}
\hypersetup{
    colorlinks=true,
    linkcolor=blue,
    citecolor=blue,
    urlcolor=blue,
}

\makeatletter
\renewcommand*\aa@pageof{, page \thepage{} of \pageref*{LastPage}}
\makeatother

\usepackage{silence}
\usepackage{natbib}
\bibpunct{(}{)}{;}{a}{}{,}

\graphicspath{
    {figures/emulators}
    {figures/mcmc}
    {figures/cosmological_parameters}
}
 \newcommand{\figext}{pdf} 
\newcommand{\includefig}[2][width=\linewidth]{\includegraphics[#1]{#2.\figext}}

\newcommand{\given}{\,\vert\,}

\newcommand{\C}{\mathcal{C}} 
\newcommand{\G}{\mathcal{G}} 
\renewcommand{\L}{\mathcal{L}} 
\newcommand{\D}{\mathcal{D}} 

\NewDocumentCommand{\rhocrit}{o}{%
  \rho_{\mathrm{cr}\IfValueT{#1}{,#1}}%
}
\newcommand{\cdm}{{\mathrm{cdm}}}

\newcommand{\ngavg}{\bar{n}_g}
\newcommand{\Mmin}{M_{\mathrm{min}}}
\newcommand{\sigmalogm}{\sigma_{\log M}}
\newcommand{\spara}{s_\parallel}
\newcommand{\rpara}{r_\parallel}
\newcommand{\sperp}{s_\perp}
\newcommand{\rperp}{r_\perp}

\newcommand{\lhat}{\hat{\ell}}
\newcommand{\lhatvec}{\vec{\lhat}}

\renewcommand{\wp}{w_p}
\newcommand{\wpofr}{\wp(\rperp)}
\newcommand{\abacus}{\textsc{Abacus}\xspace}
\newcommand{\abacusSummit}{\textsc{AbacusSummit}\xspace}
\newcommand{\abacusbase}{\textsc{AbacusBase}\xspace}
\newcommand{\abacussmall}{\textsc{AbacusSmall}\xspace}
\newcommand{\compaSO}{\textsc{CompaSO}\xspace}
\newcommand{\dq}{\textsc{DarkQuest}\xspace}
\newcommand{\emcee}{\textsc{emcee}\xspace}
\renewcommand{\sec}[1]{Sect.~\ref{#1}}

\newcommand{\fig}[1]{Fig.~\ref{#1}}

\newcommand{\tabb}[1]{Table~\ref{#1}}
\newcommand{\eq}[1]{Eq.~(\ref{#1})}
\newcommand{\eqs}[2]{Eqs.~(\ref{#1}) and (\ref{#2})}

\newcommand{\tabbcite}[1]{Table~{#1}}
\newcommand{\eqcite}[1]{Eq.~{#1}}
\newcommand{\figcite}[1]{Fig.~{#1}}
\newcommand{\munit}{\,h^{-1}M_{\odot}} 
\newcommand{\Mpc}{\,h^{-1}\mathrm{Mpc}}		
\newcommand{\Gpc}{\,h^{-1}\mathrm{Gpc}}		
\newcommand{\pclosed}[1]{\left(#1\right)}	
\newcommand{\bclosed}[1]{\left[#1\right]}	
\newcommand{\Bclosed}[1]{\left\{#1\right\}} 

\defcitealias{planckcollaborationPlanck2018Results2020}{Planck Collaboration et al. (2020)}

\begin{document}

\title{An emulation-based model for the projected correlation function}

\author{Vetle A. Vikenes,
          Cheng-Zong Ruan
          \and
          David F. Mota}

\institute{Institute of Theoretical Astrophysics, 
    University of Oslo,
    0315 Oslo, Norway\\
    \email{v.a.vikenes@astro.uio.no}
             }

\date{\today}

\abstract{Data from the ongoing \textit{Euclid} survey will map out billions of galaxies in the Universe, covering more than a third of the sky. This data will provide a wealth of information about the large-scale structure (LSS) of the Universe and will have a significant impact on cosmology in the coming years.
In this paper, we introduce an emulator-based halo model approach to forward model the relationship between cosmological parameters and the projected galaxy-galaxy two-point correlation function (2PCF). Utilizing the large \textsc{AbacusSummit} simulation suite, we emulate the 2PCF by generating mock-galaxy catalogues within the Halo Occupation Distribution (HOD) framework. 
Our emulator is designed to predict the 2PCF over scales $0.1 \leq r / (h^{-1}\text{Mpc}) \leq 105$, from which we derive the projected correlation function, independent of redshift space distortions. 
We demonstrate that the emulator accurately predicts the projected correlation function over scales $0.5 \leq r_\perp/(h^{-1}\text{Mpc}) \leq 40$, given a set of cosmological and HOD parameters. 
This model is then employed in a parameter inference analysis, showcasing its ability to constrain cosmological parameters. 
Our findings indicate that while the projected correlation function places weak constraints on several cosmological parameters due to its intrinsic lack of information, additional clustering statistics are necessary to better probe the underlying cosmology. 
Despite the simplified covariance matrix used in the likelihood model, the posterior distributions of several cosmological parameters remain broad, underscoring the need for a more comprehensive approach.

}
\maketitle

\section{Introduction}
\label{sec:intro}

    Modern galaxy redshift surveys, such as the Sloan Digital Sky Survey (SDSS; \citet{York:2000AJ....120.1579Y.SDSS}), Two Degree Field Galaxy Redshift Survey (2dFGRS; \citet{Colless:2001MNRAS.328.1039C.2dF,Cole:2005MNRAS.362..505C.2dF}), and the WiggleZ Dark Energy Survey \citep{Drinkwater:2010MNRAS.401.1429D.WiggleZ}, BOSS \citep{Dawson:2013AJ....145...10D.BOSS}, eBOSS \citep{Dawson:2016AJ....151...44D.eBOSS,eBOSSOverview:2021PhRvD.103h3533A}, DESI \citep{DESI:2016arXiv161100036D} and \textit{Euclid} \citep{laureijsEuclidDefinitionStudy2011} have enabled us to study the large-scale structure of the universe mainly traced by galaxies.
    To fully exploit the next generation of observational data, we need accurate models of galaxy clustering.

    In typical galaxy surveys, measurements of clustering statistics, such as the two-point correlation function (2PCF), are more precise at the $\mathrm{Mpc}$ scale, which is below the valid scales of the state-of-the-art perturbation theory approaches.
    The fully non-linear treatment, $N$-body simulations, is essential to accurately resolve the non-linear dynamics in structure formation.
    However, simulations are computationally demanding, posing a significant bottleneck in the exploration of cosmological parameter space.

    In contemporary galaxy formation and evolution theories, galaxies form and evolve in dark matter halos.
    After the distribution of halos is obtained from simulations, we adopt the framework of halo occupation distribution \citep[HOD;][]{Peacock:2000MNRAS.318.1144P,Seljak:2000MNRAS.318..203S.HaloModel,Berlind:2002ApJ...575..587B,zhengTheoreticalModelsHalo2005,2014MNRAS.441.2398G} as the galaxy-halo connection model.
    HOD models have been widely applied to interpret galaxy clustering data \citep[e.g.][]{Zheng:2009ApJ...707..554Z,Guo:2015MNRAS.453.4368G,Zentner:2019MNRAS.485.1196Z}. 
    In the basic form, the HOD quantifies the likelihood $P(N|M)$ that a dark matter halo of mass $M$ contains $N$ galaxies. 
    Together with the models of the spatial and velocity distributions of galaxies inside halos, the HOD framework can interpret the clustering of galaxies down to non-linear scales.

    Given the computational requirements of $N$-body simulations and advancements in machine learning techniques over the last years, \textit{emulators} have emerged as an increasingly popular tool to model galaxy clustering statistics within the HOD framework \citep[e.g.][]{kwanCosmicEmulationFast2015,kobayashiFullshapeCosmologyAnalysis2022,nishimichiDarkQuestFast2019,Zhai:2019ApJ...874...95Z,Zhai:2023ApJ...948...99Z,Kwan:2023ApJ...952...80K,cuesta-lazaroGalaxyClusteringBottom2023,cuesta-lazaroSUNBIRDSimulationbasedModel2023,Burger:2024arXiv240605098B,Storey-Fisher:2024ApJ...961..208S}. 
    The core idea of emulators is to evenly sample the parameter space and apply novel interpolation schemes to create high-accuracy estimates of clustering statistics at any point within that space.
    In this work, we build a model based on the \textsc{AbacusSummit} simulations \citep{maksimovaAbacusSummitMassiveSet2021}.

    The radial positions of galaxies are usually derived from the observed galaxy redshifts assuming a cosmological model. 
    However, these derived distances differ from the true positions due to peculiar velocities, an effect known as redshift-space distortions (RSD).
    A traditional way to probe the real-space galaxy clustering and mitigate the RSD effect is through the projected 2PCF, $w_p(r_{\perp})$, obtained by integrating the three-dimensional 2PCF $\xi^{\text{S}}(r_{\perp}, r_{\parallel})$ along the line-of-sight (LOS, $r_{\parallel}$) direction. 
    In this work, we limit our attention to the real-space galaxy 2PCF $\xi^{\text{R}}(r)$, building emulators for it and demonstrating that the emulators can recover unbiased cosmological constraints in a likelihood analysis.
    The modelling of redshift-space clustering will be presented in subsequent studies, where the emulators for $\xi^{\text{R}}(r)$ will serve as one of the model ingredients.

    The structure of this paper is as follows:
    In \sec{sec:theory} we introduce the theoretical background for the galaxy clustering. 
    In \sec{sec:constructing_mock_galaxy_catalogues} we give a brief overview of the suite of cosmological $N$-body simulations used in this work, and outline how we construct galaxy catalogues from these. 
    In \sec{sec:emulation} we present our neural network emulator.
    Then, we present our method for recovering cosmological parameters in \sec{sec:parameter_inference}, before presenting the results in \sec{sec:results}. Finally, we discuss our results in \sec{sec:discussion} before concluding in \sec{sec:conclusions}.   

\section{Theoretical background} \label{sec:theory}
The spatial distribution of galaxies can be quantified in terms of the two-point correlation function (2PCF) $\xi_{\text{gg}}(r)$, which measures the excess probability of finding a pair of galaxies at a given separation, compared to that of a random distribution. 
In cosmological surveys, the comoving distance to a galaxy, $r(z)$, is inferred from its redshift via
\begin{equation}
    r(z) = \int_0^z \frac{\dd z'}{H(z')},
\end{equation}
where $H(z)$ is the Hubble factor. We use natural units where the speed of light is $c=1$. 

In addition to the Doppler shift resulting from the Universe's expansion, light emitted from galaxies experiences an additional shift due to peculiar velocities. These peculiar velocities induce anisotropies in the measured spatial distribution of galaxies, known as redshift-space distortions (RSD). 
If a galaxy pair is separated by a vector $\Vec{r}$ in real space, it will be separated by a vector $\Vec{s}$ in redshift space. 
We adopt the \textit{distant-observer} approximation, where galaxies are assumed to be sufficiently distant such that the angle they subtend with respect to the line-of-sight (LOS) direction, denoted by $\lhatvec$, is negligible. Under this assumption, galaxies' peculiar velocity only induces a redshift in the $\lhatvec$ direction. To  the linear order in velocity, $\abs{\vec{v}}\ll1$, the redshift-space separation of galaxies due to peculiar velocities is given as
\begin{equation} \label{eq:redshift_space_real_space_relation}
    \Vec{s} = \Vec{r} + \frac{v_\parallel }{a(z)H(z)}\vec{\hat{\ell}},
\end{equation}
where $v_\parallel\equiv\vec{v}\cdot\vec{\hat{\ell}}$ is the peculiar velocity of the galaxy along the LOS, $a(z)$ is the scale factor and $H(z)$ the Hubble factor.

We decompose the separation vectors $\Vec{r}$ and $\Vec{s}$ into a component parallel and perpendicular to the LOS, $\spara,\,\rpara$ and $\sperp,\,\rperp$, respectively. Under the distant-observer approximation, RSD effects are isolated into $\spara$, meaning we have $\rperp=\sperp$. Consequently, projecting galaxies onto a plane orthogonal to $\rpara$ results in the same distribution of galaxies in real and redshift space, known as the \textit{projected correlation function},
\begin{equation} \label{eq:projected_correlation_function}
    w_p(\rperp) \equiv \int_{-\pi_\mathrm{max}}^{\pi_\mathrm{max}}\dd \spara \, \xi^{\text{S}}(\spara,\sperp) = \int_{-\pi_\mathrm{max}}^{\pi_\mathrm{max}}\dd \rpara \, \xi^{\text{R}}(r(\rpara,\rperp)), 
\end{equation}
where the equality holds if $\pi_\mathrm{max}$ is sufficiently large for correlations to be negligible beyond it. We use superscript $\text{S}$ and $\text{R}$ to denote redshift space and real space, respectively, and $r(\rpara,\rperp)\equiv \sqrt{\rpara^2 + \rperp^2}$. Importantly, the projected correlation function is independent of RSD effects under the distant-observer approximation, meaning we can model $\xi^{\text{R}}$ without having to account for peculiar velocity, which can be directly confronted with measurements from observations.

\section{Constructing mock galaxy catalogues} \label{sec:constructing_mock_galaxy_catalogues}
In this section, we briefly describe the $N$-body simulations and the catalogues of HOD galaxies used in the analyses.

\subsection{The AbacusSummit simulation suite}
\abacusSummit \citep{maksimovaAbacusSummitMassiveSet2021} is a suite of large, high-accuracy cosmological $N$-body simulations. In this work, we mainly focus on the base simulations of the suite with a volume of $2\,(h^{-1}\mathrm{Gpc})^3$ containing $6912^3$ with a mass resolution of $2\cdot 10^9\munit$. The simulation suite contains $150$ simulation boxes spanning $97$ cosmological models. They assume a flat spacetime, and the Hubble constant $H_0$ is calibrated to match the CMB acoustic scale $\theta_*$. The $8$ varying parameters are 
\begin{equation} \label{eq:cosmological_parameters}
    \C = \Bclosed{\omega_\mathrm{cdm},\omega_{\text{b}},n_s,\sigma_8,w_0,w_a,\dd n_s/\dd\ln k,N_\mathrm{eff}},
\end{equation}
where $\omega_\cdm=\Omega_\cdm h^2$ and $\omega_{\text{b}}=\Omega_{\text{b}} h^2$ are the physical cold dark matter and baryon densities, $\sigma_8$ is the amplitude of the matter power spectrum, $w_0$ is the dark energy equation of state today, $w_a$ governing the time evolution of the dark energy equation of state according to the CPL parameterization \citep{chevallierAcceleratingUniversesScaling2001,linderExploringExpansionHistory2003},
\begin{equation} \label{eq:CPL_DE_EoS}
    w = w_0 + w_a \frac{z}{1+z},
\end{equation}
where $w_0=-1$ and $w_a$ correspond to a cosmological constant,  $n_s$ is the spectral tilt, $\dd n_s/\dd\ln k$ is the running of the spectral tilt and $N_\mathrm{eff}$ is the effective number of neutrino species. 

The base simulations considered in this paper are organized into the following subsets: 
\begin{itemize}
    \item \verb|c000|: Fiducial cosmology matching Planck 2018 \citep{planckcollaborationPlanck2018Results2020} particularly the mean estimates of the TT,TE,EE+lowE+lensing likelihood chains. This cosmology contains $25$ independent realizations, referred to as the \textit{fiducial cosmology} in this paper.
    \item \verb|c001-c004|: Secondary cosmologies, including a low $\omega_\mathrm{cdm}$, thawing dark energy, high $N_\mathrm{eff}$ and low $\sigma_8$. 
    \item \verb|c100-c126|: Linear derivative grid with simulation pairs with small positive and negative steps in $8$-dimensional parameter space.
    \item \verb|c130-c181|: Broad emulator grid, covering a wider range of the full cosmological parameter space. All simulations in this grid have the same phase seed.
\end{itemize}

\tabb{tab:cosmo_params} shows the fiducial value and the prior ranges of the cosmological parameters covered by the simulations. 
\begin{table}
\caption{Cosmological parameters with their fiducial values and their prior ranges.}
\label{tab:cosmo_params}
\begin{tabular}{l c c c}
\hline
\hline
Parameter & Fiducial value & Min. prior & Max. prior \\
\hline
$N_\mathrm{eff}$ & 3.033 & 2.177 & 3.889 \\
$\dd n_s / \dd \ln k$ & 0.0 & -0.038 & 0.038 \\
$n_s$ & 0.965 & 0.901 & 1.025 \\
$\sigma_8$ & 0.808 & 0.678 & 0.938 \\
$w_0$ & -1.0 & -1.271 & -0.700 \\
$w_a$ & 0.0 & -0.628 & 0.621 \\
$\omega_b$ & 0.022 & 0.021 & 0.024 \\
$\omega_\mathrm{cdm}$ & 0.120 & 0.103 & 0.140 \\
\hline
\end{tabular}
\end{table}

The \verb|c001-c004| simulations contain $
6$ independent realizations each, but we limit ourselves to a single realization to reduce data storage requirements. In addition to the base simulations, we use the \abacussmall simulations to estimate the covariance matrix used for parameter inference analysis in \sec{sec:parameter_inference}. These contain $1687$ independent realizations of the fiducial cosmology with $1728^3$ particles in a $(0.5\Gpc)^3$ volume box. 

The group finding in each simulation is done on the fly, using a hybrid FoF-SO algorithm named \compaSO \citep{hadzhiyskaCompasoNewHalo2021}. This paper only considers halo catalogues from snapshots at a single redshift of $z=0.25$. As described in \citeauthor{hadzhiyskaCompasoNewHalo2021}, there are three types of halos in the simulation suite, and we limit our analysis to the main halo catalogues, named L1. The resulting catalogues are then populated by galaxies, following the procedure described in the following subsection.

\subsection{Modelling the galaxy-halo connection} \label{sec:modelling_the_galaxy_halo_connection}
To construct galaxy catalogues from the \abacus halos, we adopt the halo occupation distribution (HOD) model introduced by \citet{zhengTheoreticalModelsHalo2005}, where the number of galaxies in a halo is determined by the halo mass alone, meaning that we ignore galaxy assembly bias. We parameterize central galaxies according to a Bernoulli distribution and satellites according to a Poisson distribution, using functional forms that were developed to match the clustering of Luminous Red Galaxies (LRGs). 

The mean occupation number of central galaxies is parameterized as
\begin{equation} \label{eq:mean_occupation_number_central_galaxies}
    \expval{N_c}(M\,|\,\G) = \frac{1}{2}\pclosed{1+\erf\pclosed{\frac{\log M-\log \Mmin}{\sigma_\mathrm{logM}}}},
\end{equation}
where $\erf(x)$ is the error function, $\Mmin$ represents the typical halo mass required to host a central galaxy, and $\sigmalogm$ determines the width of the distribution. 

The mean occupation number of satellites is given as
\begin{equation} \label{eq:mean_occupation_number_satellite_galaxies}
    \expval{N_s}(M\,|\,\G) = \expval{N_c}(M\,|\,\G) \pclosed{\frac{M-\kappa M_\mathrm{min}}{M_1}}^\alpha,
\end{equation}
where we make the common assumption that satellite galaxies only reside in halos that already contain a central galaxy. Here, $\kappa\Mmin$ is the minimum halo mass required to host a satellite galaxy, $M_1$ can be interpreted as the typical halo mass required to host a satellite galaxy, and $\alpha$ is a power-law index governing the satellite occupation's mass dependence. 

Our HOD model comprises the five HOD parameters. However, we can fix one parameter via the mean galaxy number density
\begin{equation} \label{eq:galaxy_number_density}
    \ngavg = \int \dd M \dv{n}{M} \Big[\expval{N_c}(M\,|\,\G) + \expval{N_s}(M\,|\,\G)\Big], 
\end{equation} 
where $\dv{n}{M}$ is the halo mass function.
Given a set of $\G$, we may replace one HOD parameter with $\ngavg$ by inverting \eq{eq:galaxy_number_density}. Instead of $\Mmin$, we use $\ngavg$ as a free parameter in our model since we have strong prior information on the latter from observation data. Hence, the free model parameters considered in this work is 
\begin{equation} \label{eq:HOD_parameters}
    \G = \Bclosed{\ngavg,\sigma_\mathrm{logM},M_1,\kappa,\alpha}.
\end{equation} 

We emphasize that this HOD model is highly simplistic, and several extensions have been developed over the years to accurately incorporate more complex properties of galaxy-halo connection supported by modern observations \citep[][e.g.]{yuanABACUSHODHighlyEfficient2022}. Nonetheless, the \textit{base model} of \citet{zhengTheoreticalModelsHalo2005} in HOD-related analyses remains widely used and is well-tested, thus serving as a natural starting point to construct an emulator-based model for predicting galaxy clustering. Importantly, two-point statistics have intrinsic information loss, and we remove RSD effects by considering the projected correlation function. The base HOD model provides a benchmark for subsequent studies with more complex HOD parameterizations when additional summary statistics are included.


We follow \citet{cuesta-lazaroGalaxyClusteringBottom2023} for constructing mock galaxy catalogues, adopting $R_{200m}$ as our halo radius definition, which is the radius of a sphere where the average enclosed density is $200$ times the mean matter density of the Universe $\rho_{m0}$. The halo mass is given by the number of particles within the halo and is used to compute the halo radius. Note that our halo definitions deviate from those of \compaSO. In future analyses, we intend to fully integrate the \compaSO halo definitions to construct galaxy catalogues. Comparing results obtained from the two halo definitions will serve as a valuable test to determine the model independence of the results presented in this work, thereby ascertaining whether the underlying cosmology is indeed being probed. 

Central galaxies are located at the host halos' centre, by assumption, and satellite galaxies are distributed according to an NFW profile, $u_\mathrm{NFW}(r\vert c(M))$ \citep{navarroUniversalDensityProfile1997}. For the halo concentration, $c(M)$, we use the median concentration-mass relation from \citet{diemerAccuratePhysicalModel2019}. We assume the angular distribution of the satellite galaxies to be isotropic. 

To model RSD effects, we add velocity dispersions to the populated galaxies. By assumption, central galaxies follow their host halo and are assigned their velocity. Satellite galaxies are assigned the same velocity as the host halo, with an additional offset randomly drawn from a Gaussian distribution with mean zero and a standard deviation determined by the halo's velocity dispersion. From the velocity dispersion of galaxies we get their redshift space position according to \eq{eq:redshift_space_real_space_relation}.

\section{Emulating the galaxy two-point correlation function} \label{sec:emulation}
To predict the projected correlation function given a set of cosmological parameters, $\C$ and HOD parameters, $\G$, we construct a neural network emulator to predict $\xi^{\text{R}}(r,\C,\G)$, from which $\wp$ is obtained via \eq{eq:projected_correlation_function}. Emulators are computational models that efficiently approximate the behaviour of a system, essentially acting as high-dimensional interpolators. In this section, we describe the general setup of the neural network and its main prerequisites before introducing how mock galaxy catalogues are constructed from which the observable is computed. 

To emulate the 2PCF, we employ fully connected neural networks, which approximate a function $f$ such that 
\begin{equation}
    \vec{y} = f(\vec{x} \vert \vec{\theta}),
\end{equation}  
where $\vec{x}$ represents the features of the dataset, $\vec{y}$ the target values and $\vec{\theta}$ the trainable parameters of the network.

The activation function of the network is the Gaussian error linear unit (GELU) \citep{hendrycksGaussianErrorLinear2016}, defined as
\begin{equation} \label{eq:GELU}
    \mathrm{GELU}(x) = x\cdot\Phi(x) = x \pclosed{1 + \erf \pclosed{\frac{x}{\sqrt{2}}}},
\end{equation}
where $x$ is the output of the network's previous layer, and $\Phi(x)$ is the cumulative distribution function of the standard normal distribution. The GELU activation function is a smooth approximation of the ReLU activation function \citep{agarapDeepLearningUsing2019}. GELU exhibits more curvature and non-linearity, which may allow it to better approximate complex functions. Another advantage is that it avoids the vanishing gradient problem.

The optimal $f$ is defined by the set of parameters $\vec{\theta}$ which minimizes a loss function. In this work, we adapt the $L_1$ norm 
\begin{equation} \label{eq:L1_loss_function}
    L_1(\vec{\theta}) = \frac{1}{N} \sum_i^N \abs{y_i^\mathrm{pred}(\vec{\theta}) - y_i^\mathrm{true}},
\end{equation}
where $y_i^\mathrm{true}$ is the true value of the observable and $y_i^\mathrm{pred}(\vec{\theta})$ is the network prediction. We update the network parameters using the Adam optimizer \citep{kingmaAdamMethodStochastic2017}.

For the learning rate, we use a value of $0.01$ initially and use a learning rate scheduler to reduce the learning rate by a factor of $10$ if the validation loss does not improve after $20$ epochs until a minimum value of $10^{-6}$ is reached. This simple scheme allows for increased training efficiency in the beginning when gradients are steep, followed by increasingly finer tuning of the parameters as the loss landscape flattens. 

\subsection{Training the neural network} \label{sec:training_the_neural_network}
We train the neural network to perform the mapping
\begin{equation}
    \log_{10}\pclosed{\xi(r)} = f (\C, \G, r),
\end{equation}
where $\C$ and $\G$ are the input parameters given in \eqs{eq:cosmological_parameters}{eq:HOD_parameters}, respectively, and $r$ is the average separation distance between galaxy pairs in a given bin. Additionally, we apply standard normal scaling of the targets and features according to the transformation
\begin{equation}
    \vec{x}_i \to \frac{\vec{x}_i - \expval{\vec{x}_i}}{\sqrt{\mathrm{Var}(\vec{x}_i)}}.
\end{equation} 

To facilitate the generalization of the network, we employ weight decay and dropout. The former reduces the weights after each iteration, preventing significant emphasis from being placed on the noise of outliers, essentially working by subtracting an $L_2$ penalization term $\lambda\norm{\vec{\theta}}_2$ from \eq{eq:L1_loss_function}. Dropout counteracts co-adaption, in which each network neuron depends on specific activations of other neurons, by randomly setting neurons to $0$ with a probability $p$. Given that we have a restricted number of samples with varying cosmological parameters, dropout is an effective method to enhance generalization \citep{srivastavaDropoutSimpleWay2014}.

To further prevent overfitting, the training is stopped if the validation loss does not improve after $50$ epochs or when a maximum of $500$ epochs is reached. 

\subsection{Generating mock-galaxy catalogues} \label{sec:generating_mock-galaxy_catalogues}
Our choice of Fiducial HOD parameters is principally based on the analysis of \citet{cuesta-lazaroGalaxyClusteringBottom2023} who applied the same HOD model on the \dq simulations \citep{nishimichiDarkQuestFast2019} where the minimum halo mass is $10^{12}\munit$. In the context of these HOD parameters, \abacus halos of mass $M < 10^{12}\munit$ are improbable to host galaxies according to the occupation distributions in \eqs{eq:mean_occupation_number_central_galaxies}{eq:mean_occupation_number_satellite_galaxies}. This consideration, combined with the benefits of reduced computational cost and data storage requirements of the generated mock-galaxy catalogues, prompts us to
restrict our analysis to halos of mass $M \geq 10^{12}\munit$. There are substantial low-mass halos in the simulation suite, making them probable to contribute with additional galaxies despite low probabilities of hosting them. Our approach is overly simplistic, and an alternative approach to reducing the number of halo samples is to randomly downsample halos according to a probability distribution that depends on the halo mass, as considered by \citet[\eqcite{13}]{yuanABACUSHODHighlyEfficient2022}.

Instead of $\Mmin$, we use $\ngavg$ as a free parameter in our model since we have strong prior information on the latter. For the remaining HOD parameters, we adopt the fiducial values used by \citet[\tabbcite{3}]{cuesta-lazaroGalaxyClusteringBottom2023} and choose variations of $\pm10\%$ as our prior ranges. Our HOD parameter's fiducial values and prior ranges are shown in \tabb{tab:HOD_params}.

\begin{table}
\caption{HOD parameters with their fiducial values and their prior ranges.}
\label{tab:HOD_params}
\begin{tabular}{l c c c}
\hline
\hline
Parameter & Fiducial value & Min. prior & Max. prior \\
\hline
$\log M_1$ & 14.420 & 12.978 & 15.862 \\
$\sigma_{\log M}$ & 0.692 & 0.622 & 0.761 \\
$\kappa$ & 0.510 & 0.459 & 0.561 \\
$\alpha$ & 0.917 & 0.825 & 1.008 \\
$\log{\ngavg}$ & -3.45 & -3.7 & -3.2 \\
\hline
\end{tabular}
\tablefoot{Note: Number densities and masses are given in units of $h^3\,\mathrm{Mpc}^{-3}$ and $h^{-1}\,\mathrm{M_\odot}$, respectively.}
\end{table}

\begin{figure*}
    \centering
    \includefig[width=0.95\textwidth]{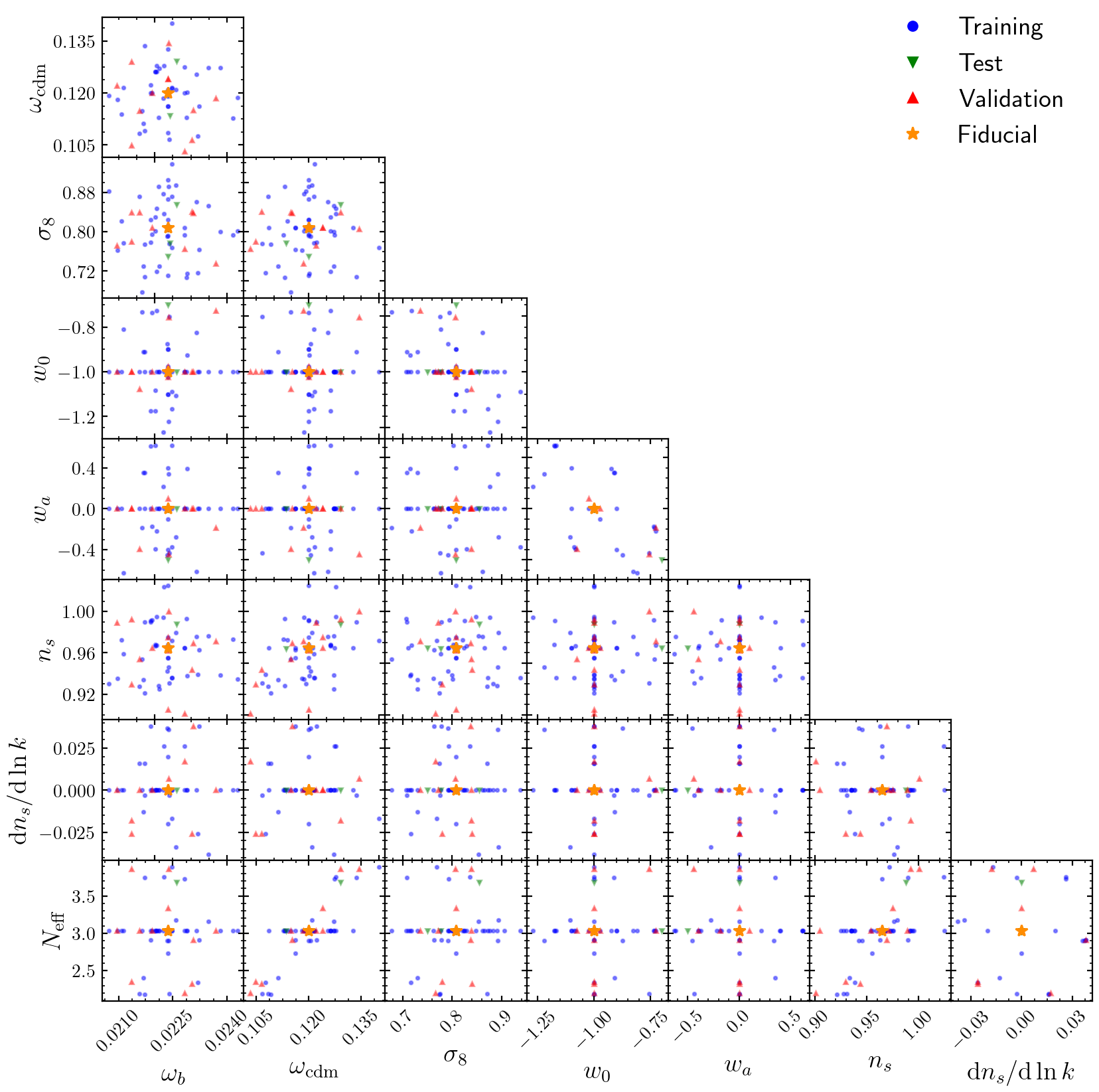}
    \caption{
        The cosmological parameters used in the training, testing and validation data sets. Note that the fiducial parameter values are included in the test data set. The large number of fiducial values for some parameters is due to the linear derivative varying two parameters at a time and the fixed parameter subsets of the broad emulator grid.
    }
    \label{fig:cosmo_params}
\end{figure*}

We randomly assign $80\%$ of the \verb|c100-c126| and \verb|c130-c180| cosmologies to the training sets and the remaining $20\%$ to the validation sets. The \verb|c000-c004| are assigned to the test sets. Degeneracies between cosmological and HOD parameters are expected, and to break these degeneracies, we construct numerous catalogues for each simulation. Breaking these degeneracies requires using numerous HOD parameter combinations for each cosmology. We therefore generate $500$ sets of HOD parameters for the training set and $100$ parameter sets each for the test and validation sets. To evenly generate samples of HOD parameters within the prior ranges in \tabb{tab:HOD_params}, we generate $5$-dimensional Latin hypercubes using the Python module \href{https://github.com/SMTorg/smt}{\textsc{SMT}} \citep{savesSMTSurrogatemodelling2024}. The cosmological parameters space covered by the \abacusbase simulations are shown in \fig{fig:cosmo_params}. 

As mentioned in \sec{sec:modelling_the_galaxy_halo_connection} we estimate the appropriate value of $\Mmin$ by inverting \eq{eq:galaxy_number_density} to generate galaxy catalogues with galaxy number densities matching the desired value. From the generated galaxy catalogues, we store the coordinates of all galaxies in real space, $(x,\,y,\,z)$, and the redshift space coordinates $s_z$ as we isolate RSD effects to the $z$-axis.


\subsection{Computing the 2PCF and its projection} \label{sec:computing_the_2PCF_and_its_projection}
We measure the real space 2PCF with the Python module {\textsc{pycorr}}\footnote{\url{https://github.com/cosmodesi/pycorr/}}, utilizing the \textsc{Corrfunc} engine \citep{sinhaCorrfuncSuiteBlazing2020,majumdarCORRFUNCBlazingFast2019}. Measurements are made using $40$ logarithmically spaced separation bin edges of $r\in[0.01,\,5)\Mpc$ and $75$ linearly spaced bin edges of $r\in[5,\,150]\Mpc$. These bin edges are chosen to cover a range of scales, ensuring a relatively high resolution on non-linear scales where we expect significant clustering and lower resolution on larger scales where galaxy pair counts are expected to be sparse. We retrieve the average separation distance between galaxy pairs in each bin from the measurements, which we use as input for the emulator.

Initial emulation tests revealed substantial errors in the predicted clustering on the largest and smallest scales, despite excluding samples with $\xi\leq10^{-7}$, as shot noise was prominent. To mitigate this, we opted to restrict the analysis to samples within the range $r\in(0.1,\,105)\Mpc$, containing only samples with $\xi^{\text{R}}(r)>10^{-7}$. 

To compute $\wpofr$ from $\xi^{\text{R}}(r)$ we use $40$ logarithmically spaced bin edges of $\rperp\in[0.5,\,40]\Mpc$. Unless otherwise stated, we use this binning for all computations of $w_p$ throughout this paper. We then create a spline of $\xi^{\text{R}}(r)$, which we integrate along $\rpara\in[0,105]\Mpc$ according to \eq{eq:projected_correlation_function}, setting $\xi^{\text{R}}=0$ for values where $r(\rperp,\rpara)$ extended beyond the training range considered.


\section{Recovering cosmological parameters} \label{sec:parameter_inference}
In this section, we demonstrate our emulator model's ability to predict cosmological parameters from a mock signal of the projected correlation function. In the future, we aim to apply our model to observational data from the ongoing \textit{Euclid} survey \citep{laureijsEuclidDefinitionStudy2011}. 
In this work, we assess the model's ability to recover cosmological parameters using mock data, which provides a controlled environment for evaluating its limitations and constraining power, setting a benchmark against unknown data. Importantly, the mock data helps us identify other summary statistics to incorporate into our model before applying it to real data.

We perform Markov Chain Monte Carlo (MCMC) parameter inference to sample the posterior distributions in our $13$-dimensional parameter space. We begin by defining the posterior distribution. Up to a constant, it is given as
\begin{equation} \label{eq:posterior_distribution_proportionality}
    p(\theta \given \D) \propto \L(\D\given\theta) p(\theta),
\end{equation}
where $\theta=\Bclosed{\C,\G}$ are the parameters we want to estimate, $\D$ is the mock data, $p(\theta\given\D)$ is the posterior of the parameters given the data and $p(\theta)$ is the prior distribution of the parameters.

The likelihood function $\L(\D\given\C,\G)$ describes how well our model, specified by our parameters, explains the observed data. We assume that the likelihood follows a multivariate Gaussian distribution, as is common when retrieving cosmological parameters from galaxy clustering observables \citep{cuesta-lazaroSUNBIRDSimulationbasedModel2023,miyatakeCosmologicalInferenceEmulator2022}, and compute the log-likelihood as
\begin{equation} \label{eq:log_likelihood_mcmc}
    \begin{split}
        \ln \L(\D\given\theta) = -\frac{1}{2} \sum_{\rperp^i,\rperp^j} &\bclosed{\wp^\mathrm{data}(\rperp^i) - \wp(\rperp^i\given\theta)} \\
        & \times \vec{C}^{-1}\pclosed{\wp^\mathrm{data}(\rperp^i), \wp^\mathrm{data}(\rperp^j)} \\
        & \times \bclosed{\wp^\mathrm{data}(\rperp^j) - \wp(\rperp^j\given\theta)} \\ 
        &+ \text{const.},
    \end{split}
\end{equation}
where $\wp^\mathrm{data}(\rperp)$ is the mock data vector, $\wp(\rperp\given\theta)$ is the emulator's prediction given the model parameters and $\vec{C}^{-1}$ is the inverse of the covariance matrix. 

\begin{figure}
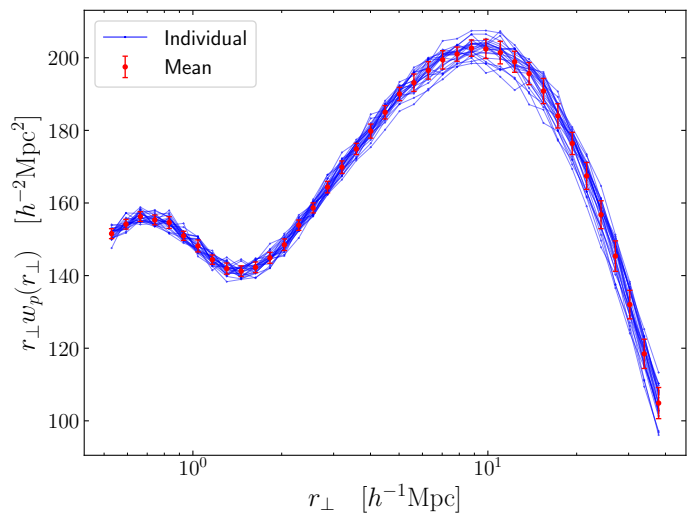

    \centering
    \includefig{wp_data_vector}
    \caption{
        Mock data vector $\wp^\mathrm{data}(\rperp)$ (red) computed from averaging over the individual $\wp$ from the $25$ fiducial cosmologies (blue). The error bars show the standard deviation of the individual $\wp$ samples.
    }
    \label{fig:wp_data_vector}
\end{figure}

We use the $25$ fiducial \verb|c000| simulation boxes to construct the mock data vector, using the fiducial HOD parameters listed in \tabb{tab:HOD_params}. We use \textsc{Corrfunc} \citep{sinhaCorrfuncSuiteBlazing2020} to estimate the projected correlation function in redshift space. The mock data signal is taken as the average $\wp(\rperp)$ from the $25$ samples, shown in \fig{fig:wp_data_vector}.

From the $25$ data vector catalogues we also compute $\xi^{\text{R}}$ over the emulator's full separation training range, $r\in(0.1,\,105)\Mpc$. The average separation vector from the $25$ samples as our fixed separation vector to predict $\xi^{\text{R}}(r\given\theta)$ during parameter inference, from which $\wp(\rperp\given\theta)$ is obtained. 

\begin{figure}
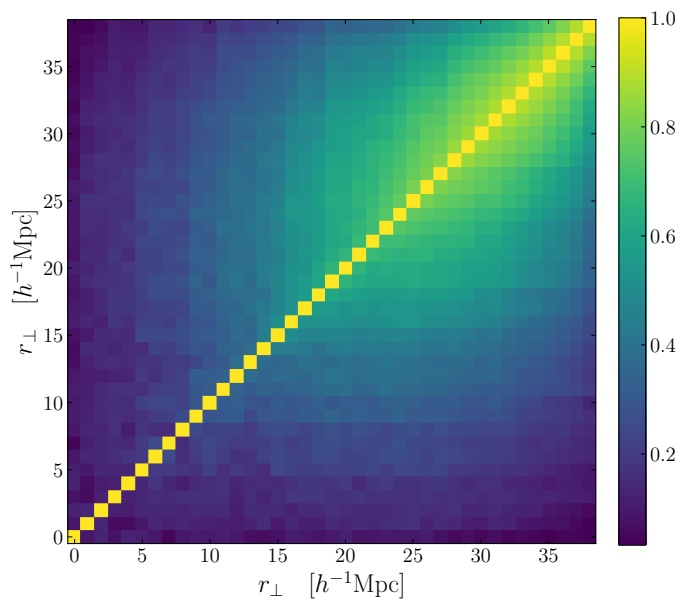

    \centering
    \includefig{corr}
    \caption{Normalized covariance matrix for $\wpofr$, computed from the $1687$ fiducial cosmology simulations in the \abacussmall data set.}
    \label{fig:covariance_matrix}
\end{figure}

We estimate the covariance matrix using the $1687$ \abacussmall simulation boxes, generating one mock-galaxy catalogue for each box using the fiducial HOD parameters in \tabb{tab:HOD_params}. A cut-off mass is not considered for these simulations, as each box contains significantly fewer halos compared to the \abacusbase boxes. The projected correlation function for each catalogue is computed according to the same prescription used to compute the mock data vector. The covariance matrix is scaled by a factor $1/64$ to match the volume of the base simulations. Figure \ref{fig:covariance_matrix} shows the normalized covariance matrix. The covariance is strictly positive, and off-diagonal terms are dominated by signals with a high signal-to-noise ratio, resulting from the large number of samples considered. The small-scale bins are primarily independent, as they are dominated by shot noise of the galaxy catalogues, whereas the large-scale bins are correlated due to cosmic variance.

Sampling is performed using the \href{https://github.com/dfm/emcee}{\emcee} python package \citep{foreman-mackeyEmceeMCMCHammer2013}, employing the differential evolution algorithm \citep{braakMarkovChainMonte2006}. The MCMC chain is assigned initial parameter values of $\C_\mathrm{fid}$ and $\G_\mathrm{fid}$, where a small Gaussian perturbation of $\mathcal{N}(0,10^{-3})$ is added to each parameter. Samples are drawn from a flat prior distribution of $\C$ and $\G$, with prior ranges given in \tabb{tab:cosmo_params} and \tabb{tab:HOD_params} respectively.

In addition to the inference analysis, we sample the posterior distribution of the cosmological parameters with HOD parameters fixed to their fiducial value and vice versa, which helps us identify and address degeneracies between the two parameter sets. This analysis is vital to identify potential supplementary statistics to add to our model to better constrain HOD parameters. Such tailored analysis is unsuitable for real data due to underestimation and bias in error measurements, underscoring the mock dataset's importance during model development.

\section{Results} \label{sec:results}
\subsection{Emulator performance} \label{subsec:emulator_results}
\begin{figure}
    \centering
    \includegraphics[width=\columnwidth]{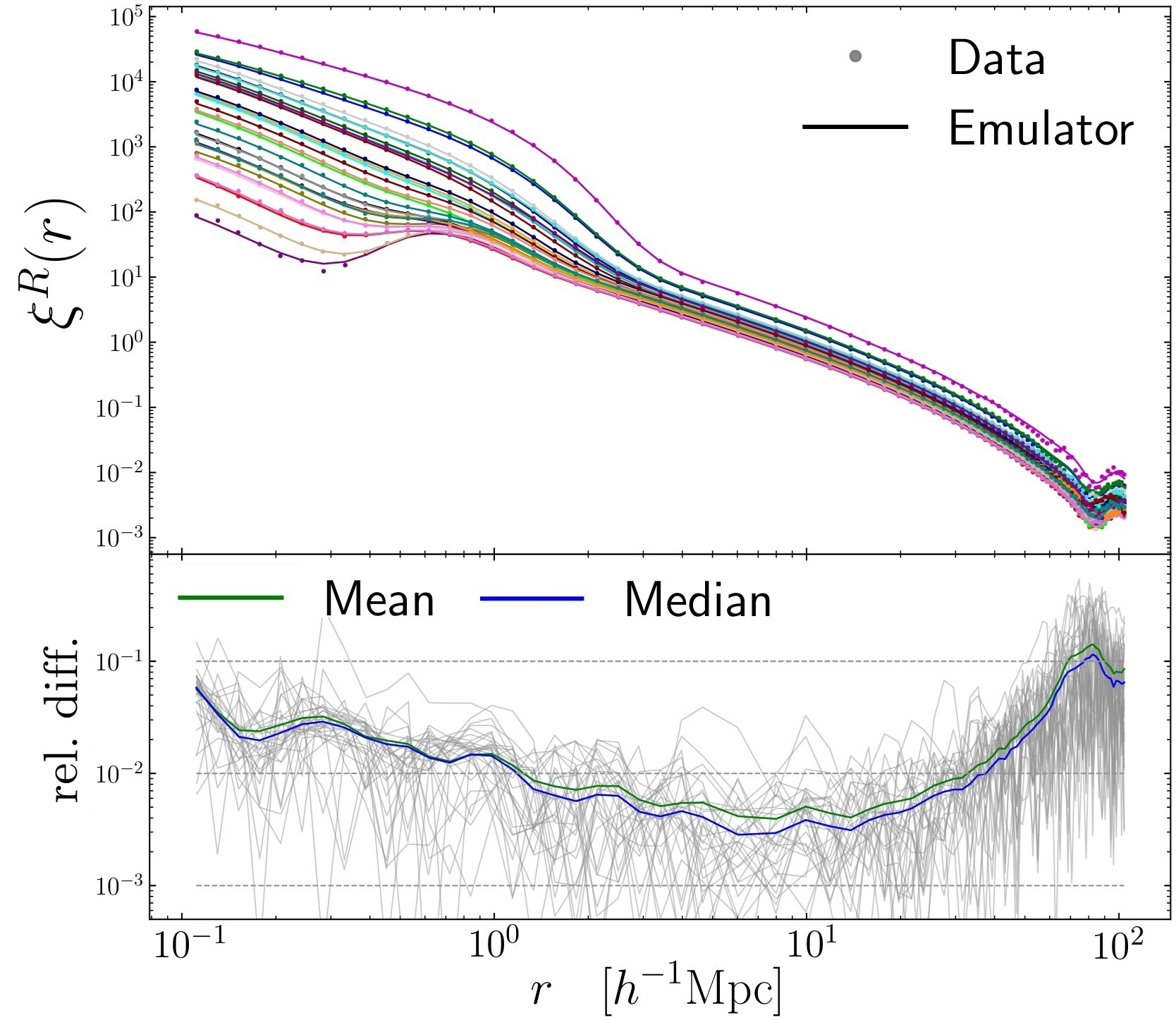}
    \caption{
        The upper panel shows one sample of $\xi^{\text{R}}$ 
        {\color{black} sampled from $29$ cosmologies in the test and validation data sets,} 
        comparing data measurements with emulator predictions. We scale $\xi^{\text{R}}$ with $r$ to emphasize the accuracy on large scales. The lower panel shows the relative error of the prediction samples from the upper panel (grey lines) and the mean and median relative error from all samples in the test data.
        }
    \label{fig:xi_emul}
\end{figure}

The emulator's prediction of $\xi^{\text{R}}$ is shown in \fig{fig:xi_emul}, with one random sample from every cosmology in the test data set shown in the upper panel. In the bottom panel, we show relative errors of the individual samples from the upper panel and include the mean and median errors computed from the test data set. The emulator accuracy increases as the extreme separation distances are approached, where Poisson noise becomes increasingly dominant. On intermediate scales, the emulator achieves sub-percent accuracy with relative errors below $1\%$ for $1\lesssim r/(\Mpc)\lesssim 30$. 

\begin{figure}
    \centering
    \includegraphics[width=\columnwidth]{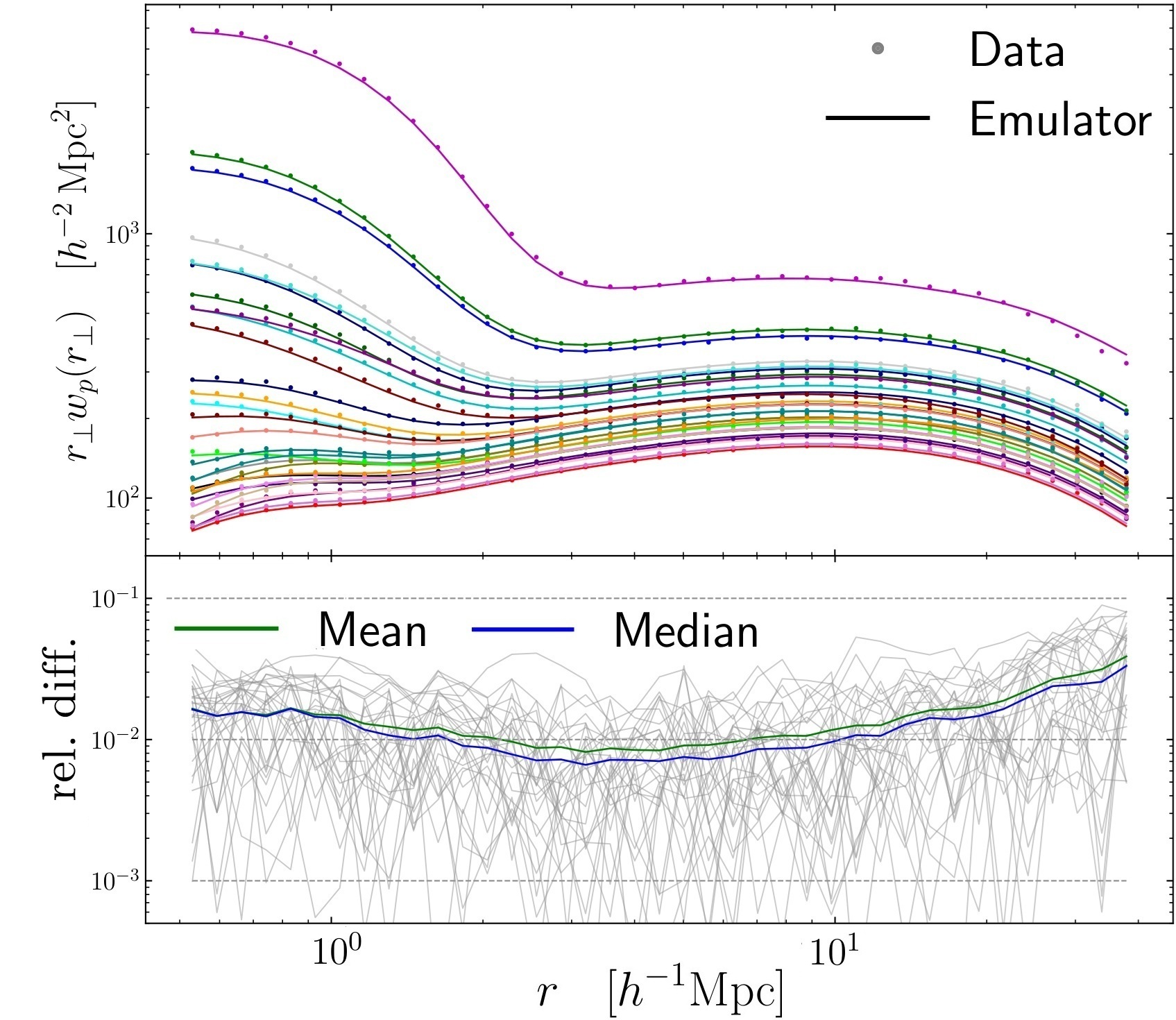}
    \caption{
        The upper panel shows one sample of $w_p$ from 
        {\color{black} sampled from $29$ cosmologies in the test and validation data sets,} 
        with solid lines corresponding to the integrated samples of $\xi^{\text{R}}$ in \fig{fig:xi_emul}. Data points show direct measurements of $\wp$ in redshift space. The lower panel shows the relative difference between $w_p$ from direct measurements and from integrating $\xi^{\text{R}}_\mathrm{pred}$.
    }
    \label{fig:wp_emul}
\end{figure}

Figure \ref{fig:wp_emul} shows the emulator prediction of $\wp$ in the upper panel, using the same test samples as in \fig{fig:xi_emul}. The data points correspond to direct measurements of $\wp$ using the redshift space coordinates of galaxies. Relative errors from each sample are shown in the lower panel, including mean and median errors computed from all test simulations. Compared to the errors of $\xi^{\text{R}}$, the emulator predicts $\wp$ more accurately, with errors below $2\%$ for $\rperp\lesssim20\Mpc$. The projection results in an overall smoothing of the relative errors of $\wp$ compared to $\xi^{\text{R}}$, with reduced errors towards the extreme ends and increased errors on intermediate scales. This indicates that some information is lost during the projection of $\xi^{\text{R}}$.

The largest emulator errors corresponded to specific HOD parameter combinations that produced noisy samples of $\xi^{\text{R}}$ on large scales where $\xi^{\text{R}}$ dropped by orders of magnitude from one measurement to the next. These outliers occurred after limiting the separation scales to $r\leq105\Mpc$. Combined with the prominent shot noise in $\xi^{\text{R}}$ at $r\gtrsim60\Mpc$, this suggests that the bin resolution was too narrow on large scales. 

We note that the \verb|c002| simulation has the highest value of $w_0$ of all cosmologies considered, and is thus outside the training range of the emulator. Nonetheless, relative error from this simulation was lower than the mean and median error from test simulations. Additionally, we did not ensure that the extreme values of the generated HOD parameters belonged to the training set, which caused some to be found in the testing and validation set.

\subsection{Parameter inference} 
\begin{figure*}
    \centering
    \includefig[width=0.8\textwidth]{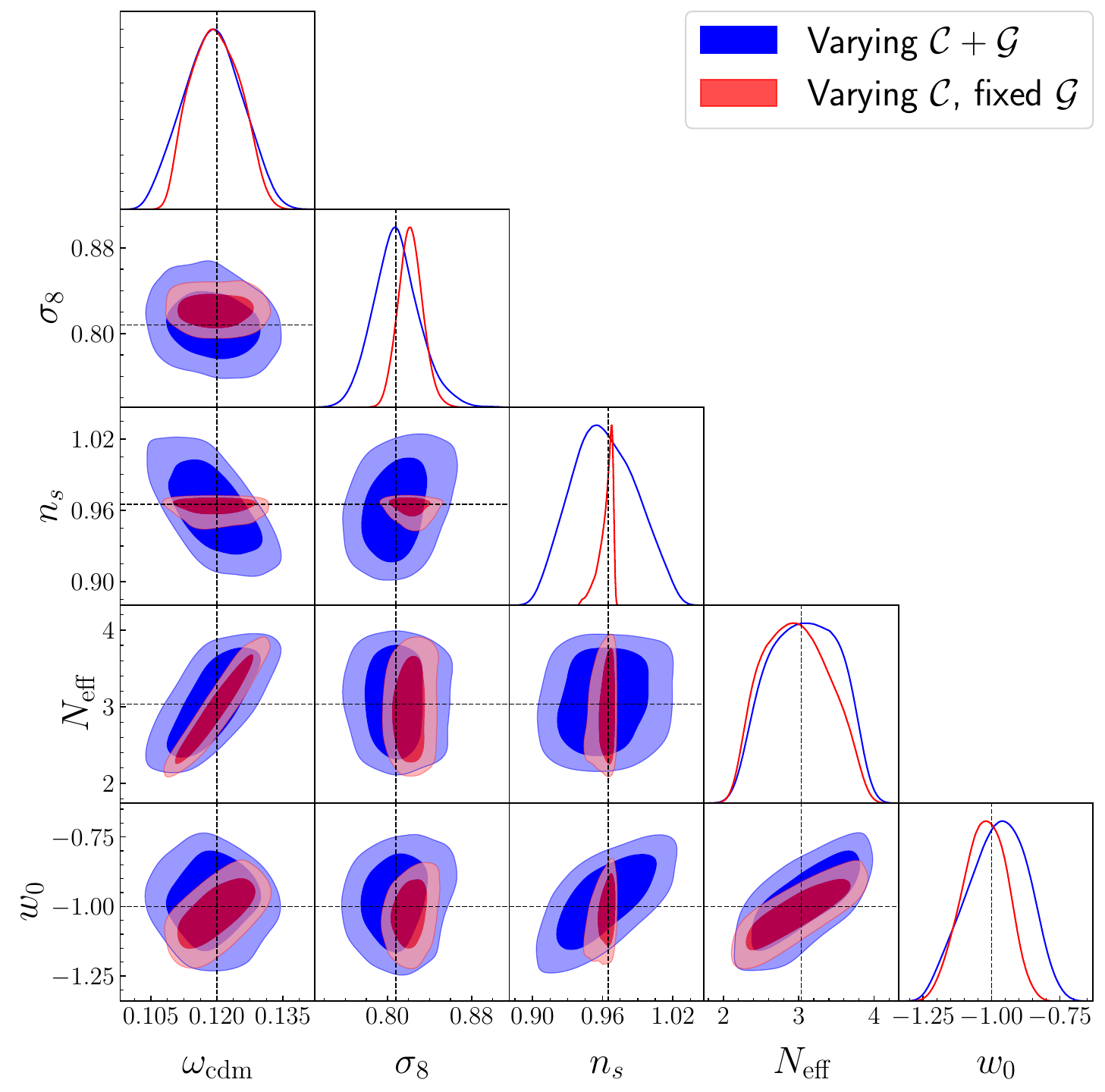}   
    \caption{
        Recovery test of the fiducial cosmology after marginalizing over the HOD parameters (in blue) and keeping the HOD parameters fixed (in red).
        }
    \label{fig:cosmo_posteriors}
\end{figure*}

Figure \ref{fig:cosmo_posteriors} shows the posterior distributions of the cosmological parameters, marginalized over the HOD parameters.
The posterior of baryonic density $\omega_b$ is poorly constrained using the projected 2PCF data alone, and therefore not shown in the plot. This is also a consequence of using an overly simplistic HOD model and the absence of the information of baryonic acoustic oscillations. Several parameters have wide marginalized distributions, indicating that our model cannot put significant constraints on them from $\wp$ at $z=0.25$ over the separation scale considered.

The posterior distributions of $\omega_\cdm$ and $\sigma_8$ suggest a tighter constraint on these parameters since the two parameters directly impact structure formation. There appears to be no considerable degeneracy between $\omega_\cdm$ and the HOD parameters. On the other hand, $\sigma_8$ shows distinct levels of degeneracy with the HOD parameters since its effect on the 2PCF can easily be compensated by changes in HOD parameters. Similar results were obtained by \citet[\figcite{9}]{miyatakeCosmologicalInferenceEmulator2022}, who were able to constrain $\Omega_m$ and $\sigma_8$ well using $\wp$ only using the \dq data.

We note that one of the EoS parameters, $w_a$, is poorly constrained by our model. 
This can be explained by the fact that our analysis is restricted to a single redshift, we cannot expect our model to appropriately account for dynamic dark energy. However, \citet[\figcite{6}]{cuesta-lazaroSUNBIRDSimulationbasedModel2023} obtains similar constraints for these parameters using $\xi^S$, but their constraints are significantly improved when including signals from density-split clustering, suggesting that the two parameters could be constrained with a single redshift only.

\begin{figure} 
    \includegraphics[width=\columnwidth]{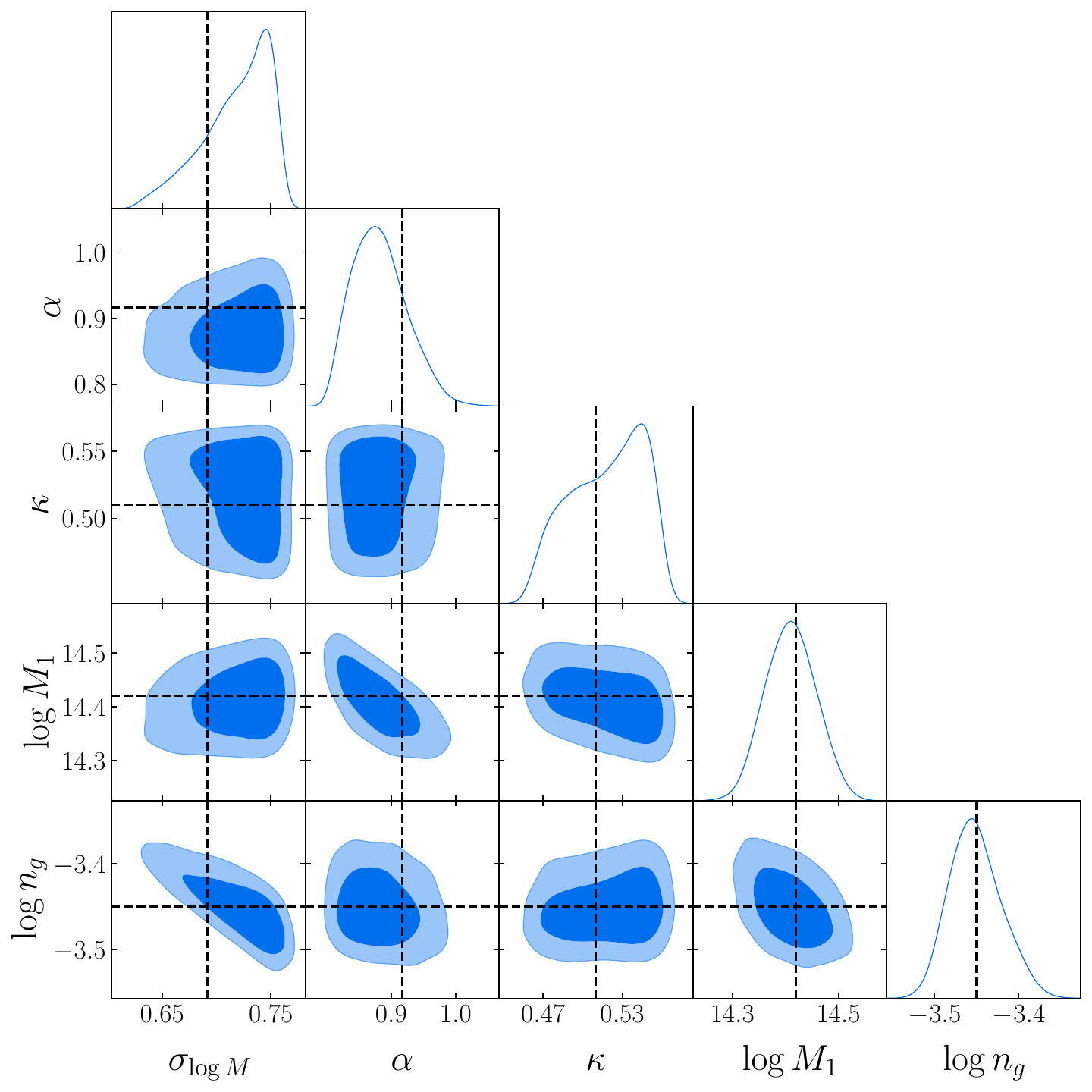}
    \caption{
        Marginalized posterior of the HOD parameters. The true parameters in the recovery test are shown in black.
        }
    \label{fig:HOD_posteriors}
\end{figure}

The HOD parameter posterior distributions are shown in \fig{fig:HOD_posteriors}. 
The HOD parameters that are well constrained are $\ngavg$ and $M_1$, which are the two parameters most strongly affecting the central and satellite galaxy distributions. Both parameters show some degeneracy with the cosmological parameters, most notably in the marginalized distributions of $\ngavg$. This is expected, as cosmological parameters affect the resulting $\ngavg$ through the HMF. The parameters $\sigma_{\log{M}}$ and $\alpha$ are relatively poorly constrained. Their posteriors, when varying HOD parameters only, suggest that wider prior ranges should be considered, as in \citet[\tabbcite{1}]{cuesta-lazaroSUNBIRDSimulationbasedModel2023}. Our results are consistent with previous studies, such as \citet[\figcite{1}]{zentnerConstraintsAssemblyBias2019}, in which HOD parameters are poorly constrained by $\wp$ alone.

\begin{figure} 
    \includegraphics[width=\columnwidth]{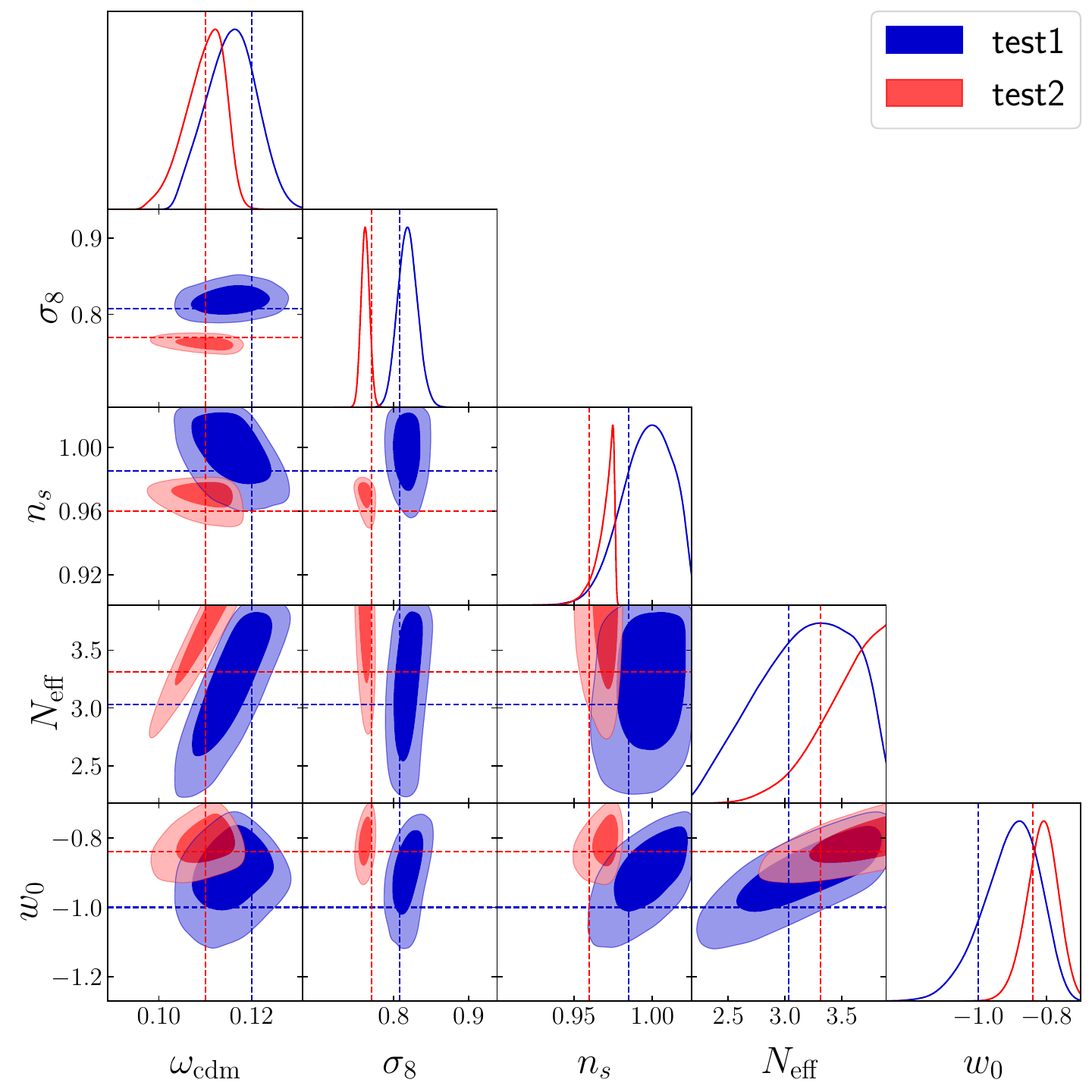}
    \caption{
        Recovery tests for two different fiducial cosmologies selected from the test data set. The true parameter values are indicated by the colored horizontal and vertical dashed lines.
        }
    \label{fig:multictest}
\end{figure}

To further validate our pipeline, we perform additional recovery tests using two different cosmologies. The results are presented in Fig.~\ref{fig:multictest}. We find that the true cosmological parameters are consistently recovered within $2\sigma$ for both cosmological models. This demonstrates the robustness of our method across different cosmologies and supports the reliability of our parameter constraining procedure.

\section{Discussion} \label{sec:discussion}

Our analysis of parameter degeneracies is mainly limited to degeneracies between the cosmological and HOD parameters. 
To accurately discern the impact on a specific parameter, it is essential to consider the intrinsic constraining power that the parameter in question exerts on the observable. 
Our analysis of these degeneracies is hindered by the model's limited ability to tightly constrain several parameters. As a result, further extensions to the model are necessary to adequately discriminate the effects of the two parameter sets. Moreover, degeneracies within each parameter set are not addressed in our current study.


A significant limitation of our recovery test is the primitive estimation of the covariance matrix. Most notably, emulator errors are not accounted for. As illustrated in \fig{fig:wp_emul}, the predictions of our emulator are less accurate at extreme scales, which is a factor that is not incorporated into the likelihood model in \eq{eq:log_likelihood_mcmc}. 
Furthermore, the errors among individual measurements of $\wp$ used to construct the mock data vector, as shown in \fig{fig:wp_data_vector}, are neglected. 
Developing a more realistic covariance matrix is the essential first step towards applying this emulator-based approach to real data. 
Consequently, we are unable to draw definitive conclusions about the true constraining power of our model.

After refining the current likelihood estimation of our model, a second clustering statistic should be added before further explorations of the information content of our inference analysis is considered, such as the scale dependence of parameter's constraints. 

Bias parameters are among the critical ingredients missing in the base model of \citet{zhengTheoreticalModelsHalo2005}. Both \textit{halo assembly bias}, referring to halos of equal mass exhibiting different clustering properties, and \textit{galaxy assembly bias}, referring to variations in the number of galaxies hosted by halos of identical mass, have been omitted in our analysis. \citet[\figcite{10}]{cuesta-lazaroGalaxyClusteringBottom2023}, for instance, shows how including environmental-based assembly bias can induce a shift in the constraint of $\sigma_8$ more than $1\sigma$ from their fiducial value. Other effects, such as velocity biases, are often introduced to allow galaxies' peculiar velocity to be slightly offset from that of their host halo, which is supported by observational evidence \citep{yuanABACUSHODHighlyEfficient2022}. In future analyses, we plan to construct refined galaxy catalogues using the highly efficient \textsc{Python} package \textsc{AbacusHOD} \citep{yuanABACUSHODHighlyEfficient2022} which includes a large variety of modern HOD variations and is specifically designed to work with the \abacusSummit simulation suite.  

{\color{black} 
Several works in the literature also emulate galaxy clustering statistics using different simulation suites, such as the \textsc{Aemulus} project \citep{Zhai:2019ApJ...874...95Z,Zhai:2023ApJ...948...99Z,Storey-Fisher:2024ApJ...961..208S} and the Mira-Titan project \citep{kwanGalaxyClusteringMiraTitan2023}. These simulations cover a similar range of cosmological parameter space as the \abacus simulations, but feature different extensions to the baseline $\Lambda$CDM model, such as massive neutrinos and dynamical dark energy. In contrast to previous works, which typically use Gaussian Process (GP) regression, we adopt fully-connected neural networks for our emulator construction. Neural networks are particularly well-suited for handling large datasets, unlike GP regression, which scales as $\mathcal{O}(N^3)$ with $N$ being the size of the training dataset.
}

\section{Conclusions} \label{sec:conclusions}

Emulators provide an efficient method to bypass the prohibitively expensive computational demands of running full $N$-body simulations for modelling galaxy clustering statistics on non-linear scales. 
In this paper, we have introduced a model designed to predict the real-space correlation function of galaxies over the range $r\in[0.1,\,105]\Mpc$.
The key strength of this approach lies in its computational efficiency, allowing for near-instantaneous estimation of the correlation function given a set of cosmological parameters. This was achieved by constructing galaxy catalogues using a straightforward Halo Occupation Distribution (HOD) model applied to halos from the \textit{AbacusSummit} simulation suite.

Our emulator predicts the projected correlation function over a wide range of scales, with sub-percent accuracies on scales $1 \lesssim \rperp/(\Mpc) \lesssim 10$. Using this model, we conducted inference analyses to constrain key cosmological parameters to constrain key cosmological parameters, particularly $\omega_\mathrm{cdm}$ and $\sigma_8$.
While these analyses demonstrate the model's capability, the limited information provided by the projected correlation function alone restricts us to place stronger constraints on other cosmological parameters. 
This issue can be addressed by incorporating the information from redshift-space clustering, which we plan to explore in future work.
Furthermore, our analysis reveals that the simplified covariance matrix employed does not account for emulator errors, highlighting the need for further refinement of our methodology before drawing definitive conclusions. Despite these limitations, our results are consistent with previous studies regarding the constraining power of the projected correlation function.

{\color{black}Future work in developing an emulator-based model for large-scale structure analysis will focus on incorporating additional summary statistics to enhance the model's ability to probe underlying cosmological information, such as higher-order clustering statistics, density-split clustering \citep{cuesta-lazaroSUNBIRDSimulationbasedModel2023,Paillas:2024MNRAS.531..898P} and the marked correlation function \citep{Storey-Fisher:2024ApJ...961..208S}}. Additionally, by employing a more sophisticated HOD model, we aim to generate galaxy catalogues that more accurately reflect the complexities of galaxy clustering. At present, our model cannot robustly marginalize over the HOD parameters, a limitation we intend to address. Once this is achieved, we plan to apply our model to observational data from the ongoing \textit{Euclid} survey to further refine cosmological parameter estimates.


\section*{Acknowledgements}
We thank the Research Council of Norway for their support and the resources provided by UNINETT Sigma2 -- the National Infrastructure for High-Performance Computing and Data Storage in Norway. 
We are grateful to the editor and the anonymous reviewer for their valuable comments and suggestions.

\clearpage 
\bibliographystyle{aa}

\bibliography{ProjCorrFuncArticle}



\end{document}